\documentclass[fleqn,pra,twocolumn]{revtex4}

\usepackage{amsmath,graphicx}
\usepackage{dcolumn}
\usepackage{verbatim}
\usepackage{appendix}

\begin{document}

\title{High-precision solution of the Dirac Equation for \\
the hydrogen molecular ion using a basis-set expansion}

\author{Hugo D. Nogueira$^1$ and Jean-Philippe Karr$^{1,2}$}
\affiliation{$^1$Laboratoire Kastler Brossel, Sorbonne Universit\'e, CNRS, ENS-Universit\'e PSL, Coll\`ege de France, 4 place Jussieu, F-75005 Paris, France}
\affiliation{$^2$Universit\'e d'Evry-Val d'Essonne, Universit\'e Paris-Saclay, Boulevard Fran\c cois Mitterrand, F-91000 Evry, France}

\begin{abstract}
The Dirac equation for H$_2^+$ is solved numerically by expansion in a basis set of two-center exponential functions, using different kinetic balance schemes. Very high precision (27-32 digits) is achieved, either with the dual kinetic balance, which provides the fastest convergence, or without imposing any kinetic balance condition. Application to heavy molecular ions is also illustrated. Calculation of relativistic sum rules shows that this method gives an accurate representation of the complete Dirac spectrum, making it a promising tool for calculations of QED corrections in molecular systems.
\end{abstract}

\maketitle

\section{Introduction} \label{sec-intro}

The relativistic two-center Coulomb problem plays a fundamental role in molecular physics, similarly to the one-center problem in atomic physics. It is also of interest for applications in two distinct regimes. Firstly, the lightest molecular ions H$_2^+$, HD$^+$, etc., are studied experimentally~\cite{Alighanbari20,Patra20,Kortunov21} and theoretically~\cite{Korobov17,Korobov21} with high precision. A precise solution of the two-center Dirac equation can be used to develop the theory further through nonpertubative calculations of QED corrections, in order to improve determinations of fundamental constants~\cite{Karr22} and constraints on additional forces beyond the Standard Model~\cite{Alighanbari20,Germann21,Delaunay22}. Secondly, in the strong-field regime, quasi-molecules formed in collisions of highly charged heavy ions are unique tools to explore phenomena related to the instability of the QED vacuum~\cite{Greiner85,Maltsev19}. Precise energy level calculations in these systems, including QED corrections~\cite{Artemyev15}, are useful to guide experimental efforts.

Substantial progress in the numerical resolution of the two-center Dirac equation has been achieved recently. The relativistic energy of H$_2^+$ was calculated with 20-digit accuracy in two independent works, one by Kullie and Schiller using the finite element method~\cite{Kullie22}, and our previous work~\cite{Nogueira22} using an iterative method~\cite{Kutzelnigg89}.

Although it would be natural to think that those results are already sufficient, there is actually a strong interest in going even further in the perspective of performing nonperturbative calculations of QED corrections in the hydrogen molecular ions, in particular the one-loop self-energy, which is currently the main source of theoretical uncertainty~\cite{Korobov17,Korobov21}. Indeed, the calculation of the one-loop self-energy in a weak binding field (i.e. low nuclear charges) suffers from a serious loss of numerical precision because of strong cancellations occurring in the renormalization procedure, hence the need for extremely accurate wavefunctions and energies~\cite{Jentschura01}.

Furthermore, calculations of QED corrections require knowledge of the Dirac Green function, thus a numerical representation of the whole spectrum of the Dirac Hamiltonian. A numerical method that provides only a single eigenstate at each execution, such as those presented in~\cite{Kullie22,Nogueira22}, might prove impractical for this purpose, and it is more desirable to use an expansion of the wavefunctions in a finite basis set. The Dirac equation is then transformed into an eigenvalue problem that can be fully diagonalized, allowing for numerical evaluation of the Green function~\cite{Yerokhin20}.

Various types of basis sets have been used to expand the four-component Dirac wavefunction of the two-center problem~\cite{Pavlik67,Mueller76,Mark87,LaJohn92,Parpia95,Franke97,Artemyev10,Fillion12,Tupitsyn14}, such as Gaussians~\cite{Mark87,Parpia95,Franke97}, Slater orbitals~\cite{LaJohn92}, or B-splines~\cite{Artemyev10,Fillion12}. In this work, we use a basis set of pure two-center exponentials~\cite{Tsogbayar06}, similar to that used in our previous work~\cite{Nogueira22}. Compared to e.g. a Gaussian basis set, it allows for a better representation of the singular behavior of the wavefunction in the vicinity of the (point-like) nuclei.

One of the main issues encountered when solving the Dirac Equation in a basis expansion is the so-called variational collapse, which manifests itself by the presence of spurious states in the forbidden region between the lowest positive-energy eigenvalue and the highest negative-energy one~\cite{Kutzelnigg84,Lewin13}. Several strategies have been developed to avoid this problem~\cite{Kutzelnigg84,Lewin13,Talman86,LaJohn92,Dolbeault03,Hill94,Stanton84,Dyall90,Visscher91,Shabaev04}. One of them is the use of a min-max variational principle~\cite{Talman86,Dolbeault03,LaJohn92}, which, however, necessitates the resolution of a computationally expensive nonlinear eigenvalue problem. The most widely used approach is the kinetic balance, whereby some relationship between the spinor components of the basis functions is imposed. The earliest and most popular version of this idea is the restricted kinetic balance (RKB)~\cite{Stanton84,Dyall90}, which guarantees that the kinetic energy for positive-energy states is correct in the non-relativistic limit. An important refinement is the dual kinetic balance (DKB)~\cite{Shabaev04,Kotov21} that was shown to avoid spurious states in the central field case. In the DKB, positive- and negative-energy states are described on an equal footing, which is a favorable feature for evaluation of QED corrections that involve sums over the whole spectrum. Nevertheless, a rigorous mathematical study of the spurious state problem showed that their absence is not fully guaranteed in any of the above approaches for a pure Coulomb attractive potential~\cite{Lewin13}. On the other hand, their presence is not an insurmountable problem in practical calculations as they can be identified and eliminated~\cite{Drake81,Artemyev10}.

In this work, we investigate two different approaches. Firstly, we solve the Dirac Equation in the absence of any kinetic balance condition, an approach we shall call ``no kinetic balance'' (NKB), and secondly, we use a DKB basis set~\cite{Shabaev04}. In addition, we performed calculations using the RKB, which are described in the Appendix~\ref{sec-app-rkb}. By analyzing the convergence of our numerical results, we show that both the NKB and DKB approaches implemented with an exponential basis set improve the accuracy of relativistic energy levels by several orders of magnitude with respect to~\cite{Kullie22,Nogueira22}.

In the perspective of performing nonperturbative QED calculations, it is not sufficient to study the precision of the energy of the ground state or first few excited states; one should rather assess to which extent the discrete spectrum obtained by diagonalizing the Dirac Hamiltonian in a finite basis set represents its actual spectrum. To this end, we evaluate several sum rules~\cite{Drake81}, demonstrating the suitability of our approach to construct a numerical approximation of the Dirac Green function.

\section{Matrix representation of the Dirac equation} \label{sec-dirac-eq}

We write the Dirac equation in atomic units ($\hbar = m = e = 1$) as
\begin{subequations} \label{eq_dirac}
\begin{flalign}
&H_D \psi = E \psi, \;\;\; \psi = 
    \begin{pmatrix}
    \varphi \\
    \chi
    \end{pmatrix},& \\
&H_D = (\beta\!-\!I_4) c^2 + c \boldsymbol{\alpha}\mathbf{p} + V = 
    \begin{pmatrix}
      V & c \boldsymbol{\sigma}\mathbf{p} \\
      c \boldsymbol{\sigma}\mathbf{p}  & V \!-\! 2 c^2
    \end{pmatrix},&
\end{flalign}
\end{subequations}
where $H_D$ is the Dirac Hamiltonian, and $\psi$ is the four-component wavefunction, composed by the large, $\varphi$, and small, $\chi$, components. In Eq.~(\ref{eq_dirac}b), $\beta$ and $\boldsymbol{\alpha}$ are the Dirac matrices, $\boldsymbol{\sigma}$ the Pauli matrices, and $I_4$ is the $4\times 4$ identity matrix. The Coulomb potential $V$ is given by
\begin{equation} \label{eq_V}
V = -\frac{Z_1}{r_1} - \frac{Z_2}{r_2} \,,
\end{equation}
where $Z_1$, $Z_2$ are the nuclear charges and $r_1$, $r_2$ the distances from each nucleus to the electron. Note that the rest mass energy $c^2$ has been subtracted from the energy in Eq.~(\ref{eq_dirac}b).

The large (small) components of the wavefunctions can be expanded in a basis set $\left \{g_{\mu} \right\}$ ($\left \{f_{\mu} \right\}$) with linear coefficients $A_{\mu}$ ($B_{\mu}$):
\begin{equation}
\varphi = \sum_{\mu = 1}^{2N}  A_{\mu} g_{\mu} \, , \;\; \chi = \sum_{\mu = 1}^{2N} B_{\mu} f_{\mu} \,,
\label{eq-basis-exp}
\end{equation}
where the functions $f_{\mu}$ and $g_{\mu}$ have opposite parities. Here, we have adopted notations similar to those of Ref.~\cite{Sun11}. $\varphi$ and $\chi$ comprise two spinor components, so that $N$ is the number of terms in the expansion for a single spinor component. 

Kinetic balance conditions consist in imposing some relationship between the basis functions $g_{\mu}$ and $f_{\mu}$. The matrix representation of the Dirac equation depends on the chosen condition. Following~\cite{Sun11}, we give below this representation for the NKB and DKB schemes. Expressions for the RKB are given in the Appendix~\ref{sec-app-rkb}.

\subsection{No kinetic balance}

With the ansatz~(\ref{eq-basis-exp}), the Dirac equation~(\ref{eq_dirac}) writes, in matrix form,
\begin{equation}\label{eq_nkb}
\begin{pmatrix}
\mathbf{V}^{LL} & c\boldsymbol{\Pi}^{LS} \\
c\boldsymbol{\Pi}^{SL} & \mathbf{V}^{SS}\!-\!2c^2 \mathbf{S}^{SS}
\end{pmatrix}
\begin{pmatrix}
\mathbf{A} \\
\mathbf{B}
\end{pmatrix}
 = E \begin{pmatrix}
\mathbf{S}^{LL} & 0 \\
0   & \mathbf{S}^{SS}
\end{pmatrix}
\begin{pmatrix}
\mathbf{A} \\
\mathbf{B}
\end{pmatrix}
\end{equation}
where the matrix elements are given by
\begin{equation} \label{eq_elem_nkb}
\begin{aligned}
 & \mathbf{V}_{\mu\nu}^{LL}=\left\langle g_{\mu}|V|g_{\nu}\right\rangle,
 \mathbf{V}_{\mu\nu}^{SS}=\left\langle f_{\mu}|V|f_{\nu}\right\rangle, \boldsymbol{\Pi}_{\mu\nu}^{LS}=\left\langle g_{\mu}|\boldsymbol{\sigma}\textbf{\text{p}}|f_{\nu}\right\rangle,\\
 & \boldsymbol{\Pi}_{\mu\nu}^{SL}=\left\langle f_{\mu}|\boldsymbol{\sigma}\textbf{\text{p}}|g_{\nu}\right\rangle, \mathbf{S}_{\mu\nu}^{LL}=\left\langle g_{\mu}|g_{\nu}\right\rangle,
  \mathbf{S}_{\mu\nu}^{SS}=\left\langle f_{\mu}|f_{\nu}\right\rangle.
\end{aligned}
\end{equation}

\subsection{Dual kinetic balance}

The DKB combines the RKB (see Eq.~(\ref{eq_basis_rkb})) and ``inverse kinetic balance''~\cite{Sun11} prescriptions to ensure correct description of both positive- and negative-energy states in the nonrelativistic limit. The wavefunction is expanded as 
\begin{equation} \label{eq_basis_dkb}
\begin{pmatrix}
\varphi \\
\chi
\end{pmatrix} = \sum_{\mu=1}^{2N} \left[
A_{\mu}
\begin{pmatrix}
g_{\mu} \\
\frac{1}{2c} \boldsymbol{\sigma}\mathbf{p} \, g_{\mu}
\end{pmatrix}
+
B_{\mu}
\begin{pmatrix}
-\frac{1}{2c} \boldsymbol{\sigma}\mathbf{p} \, f_{\mu} \\
f_{\mu}
\end{pmatrix}
\right] .
\end{equation}
The Dirac equation is then written in matrix form as
\begin{equation} \label{eq_dkb}
\begin{aligned}
&\,\begin{pmatrix}
\mathbf{T}^{LL} + \mathbf{V}^{LL} + \frac{1}{4c^2}\mathbf{W}^{LL} & c\boldsymbol{W}^{LS} \\
c\boldsymbol{\Pi}^{SL} & \mathbf{V}^{SS} -2c^2 \mathbf{S}^{SS}
\end{pmatrix}
\begin{pmatrix}
\mathbf{A} \\
\mathbf{B}
\end{pmatrix} \\
& = E \begin{pmatrix}
\mathbf{S}^{LL} + \frac{1}{2c^2}\mathbf{T}^{LL} & 0 \\
0   & \mathbf{S}^{SS} + \frac{1}{2c^2}\mathbf{T}^{SS}
\end{pmatrix}
\begin{pmatrix}
\mathbf{A} \\
\mathbf{B}
\end{pmatrix},
\end{aligned}
\end{equation}
where the matrix elements are given by
\begin{equation} \label{eq_elem_dkb}
\begin{aligned}
   &\,\mathbf{T}_{\mu\nu}^{LL}=\left\langle g_{\mu}\left|p^2/2\right|g_{\nu}\right\rangle,
   \mathbf{V}_{\mu\nu}^{LL}=\left\langle g_{\mu}|V|g_{\nu}\right\rangle, \\
   & \mathbf{W}_{\mu\nu}^{LL}=\left\langle g_{\mu}|\boldsymbol{\sigma}\textbf{\text{p}}V\boldsymbol{\sigma}\textbf{\text{p}}|g_{\nu}\right\rangle, 
   \mathbf{S}_{\mu\nu}^{LL}=\left\langle g_{\mu}|g_{\nu}\right\rangle, \\
   & \mathbf{W}_{\mu\nu}^{LS}=\left\langle g_{\mu}|\boldsymbol{\sigma}\textbf{\text{p}}V - V\boldsymbol{\sigma}\textbf{\text{p}} - T\boldsymbol{\sigma}\textbf{\text{p}}|f_{\nu}\right\rangle, \\
   & \mathbf{W}_{\mu\nu}^{SL}=\left\langle g_{\mu}|V\boldsymbol{\sigma}\textbf{\text{p}} - \boldsymbol{\sigma}\textbf{\text{p}}V - \boldsymbol{\sigma}\textbf{\text{p}}T|f_{\nu}\right\rangle, \\
   & \mathbf{T}_{\mu\nu}^{SS}=\left\langle f_{\mu}|V|f_{\nu}\right\rangle,
   \mathbf{V}_{\mu\nu}^{SS}=\left\langle f_{\mu}|V|f_{\nu}\right\rangle, \\
   & \mathbf{W}_{\mu\nu}^{SS}=\left\langle f_{\mu}|\boldsymbol{\sigma}\textbf{\text{p}}V\boldsymbol{\sigma}\textbf{\text{p}}|f_{\nu}\right\rangle, 
   \mathbf{S}_{\mu\nu}^{SS}=\left\langle f_{\mu}|f_{\nu}\right\rangle .
\end{aligned}
\end{equation}

\section{Real exponential basis set and numerical details} \label{sec-method}

We use a basis set of real exponential functions~\cite{Tsogbayar06,Nogueira22}:
\begin{equation} 
g_{\mu}^{(i)} (\mathbf{r}) = e^{im^{(i)}\phi} r^{|m^{(i)}|} \left( e^{-\alpha_{\mu} r_1 -\beta_{\mu} r_2} \pm e^{-\beta_{\mu} r_1 -\alpha_{\mu} r_2} \right), \label{eq_basis_function}
\end{equation}
with $\mu = 1, \ldots, N$. The index $i = 1,2$ represents the spinor component; the projection of the spin ($\mathbf{s}$) on the internuclear axis $z$ is $s_z = 1/2$ ($-1/2$) for $i=1(2)$. $\phi$ is the angle of rotation around $z$, and $r$ the distance from this axis to the electron. $m$ is an eigenvalue of $l_z$, $\mathbf{l}$ being the orbital momentum. For example, for a state of $j_z = 1/2$ ($\mathbf{j} = \mathbf{l} + \mathbf{s}$), $m$ takes on the value $0$ for $i = 1$ and $1$ for $i = 2$. The sign in the right-hand side is equal to $(-1)^{m^{(j)}}$ for \textit{gerade} states and $-(-1)^{m^{(j)}}$ for \textit{ungerade} states. The basis functions for the small components $f_{\mu}$ are identical to $g_{\mu}$, except for the fact that they are of opposite parity.

The exponents $\alpha_{\mu}$ and $\beta_{\mu}$ are chosen in a pseudorandom way~\cite{Tsogbayar06} in several intervals, see Table~I of~\cite{Nogueira22} for an illustrative example. The first three intervals comprise smaller values of the exponents ($\alpha_{\mu}, \beta_{\mu} \sim 1$) and mainly influence the behaviour of the wavefunctions at intermediate ($r_1,r_2 \sim a_0$, where $a_0$ is Bohr's radius) and long distances. The other intervals including increasingly large exponents model their singular behaviour in the vicinity of the nuclei. In contradistinction with~\cite{Nogueira22}, the sizes $n_i$ of all the subsets are here chosen to be equal.

One important advantage of this basis set is to better represent the singular behavior of the wavefunction in the vicinity of the point-like nuclei (through the inclusion of large exponents in the basis) compared to, e.g., Gaussians. Moreover, all the matrix elements appearing in Eqs.~(\ref{eq_elem_nkb}), and (\ref{eq_elem_dkb}) can be calculated analytically by recurrence relations (see~\cite{Tsogbayar06}) which allows for a high level of accuracy, as numerical integrations are completely avoided.

Since we aim for highly accurate energy levels and wavefunctions, it is mandatory to use multiple-precision arithmetic. The very wide range of exponents included in basis sets makes the matrices ill-conditioned and increases further the need for numerical acccuracy. Multi-precision arithmetic is handled by the package MPFUN2020~\cite{Bailey23}. For most calculations we use $96$-digit floating point numbers. We checked the stability of our results as a function of numerical precision; in cases where a non-negligible dependence was observed, the numerical precision was increased so that all given digits are stable.

The calculation of matrix elements is much more computationally expensive for DKB than for NKB (compare Eqs.~(\ref{eq_elem_nkb}) and (\ref{eq_elem_dkb})). For example, for $N = 1000$ and $96$-digit arithmetic, it required about half an hour in NKB and 19 hours in DKB, using 12 cores of an Intel Xeon Gold 5220 processor.

\section{Results} \label{sec-results}

Unless otherwise specified, we use the CODATA 2018 value of $c = \alpha^{-1}$, $c_a =137.035\,999\,084$, in all calculations~\cite{Tiesinga21}.

\begin{figure}[t]
\centering
\includegraphics[width=\columnwidth]{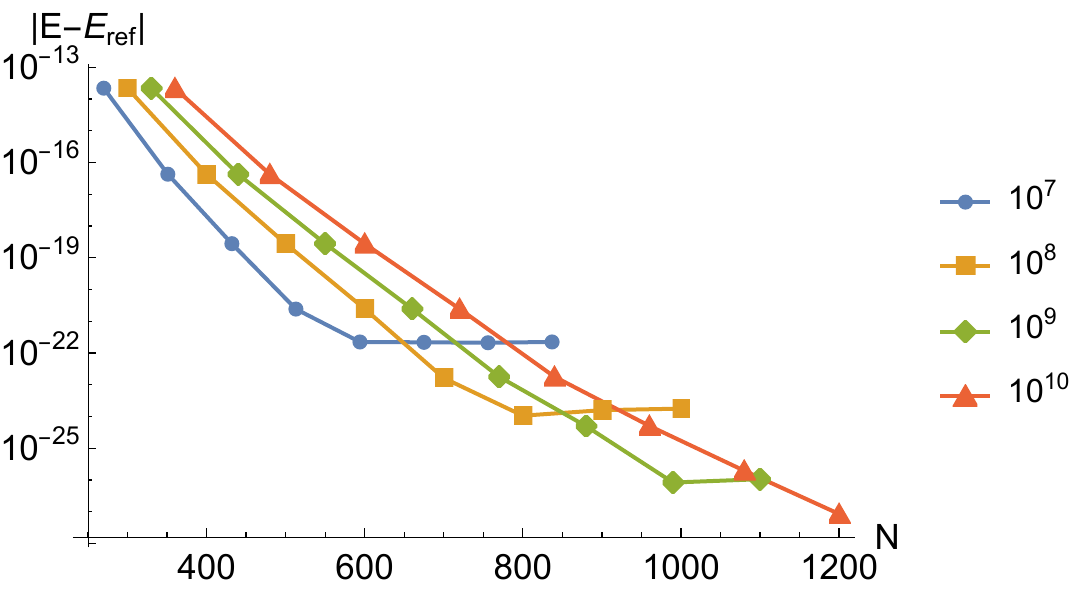}\\
\caption{Convergence of the ground-state energy of H$_2^+$ at $R = 2.0$ with the NKB basis set, using different values of the maximal exponent included in the basis set, $\alpha_{\rm max}$, which are given in the legend. The reference value of the energy, $E_{\rm ref}$, used to estimate the error $|E - E_{\rm ref}|$ is given in the last line of Table~\ref{tab_r_dkb}.}
\label{conv_e_nkb}
\end{figure}

\begin{figure}[h]
\centering
  \includegraphics[width=\columnwidth]{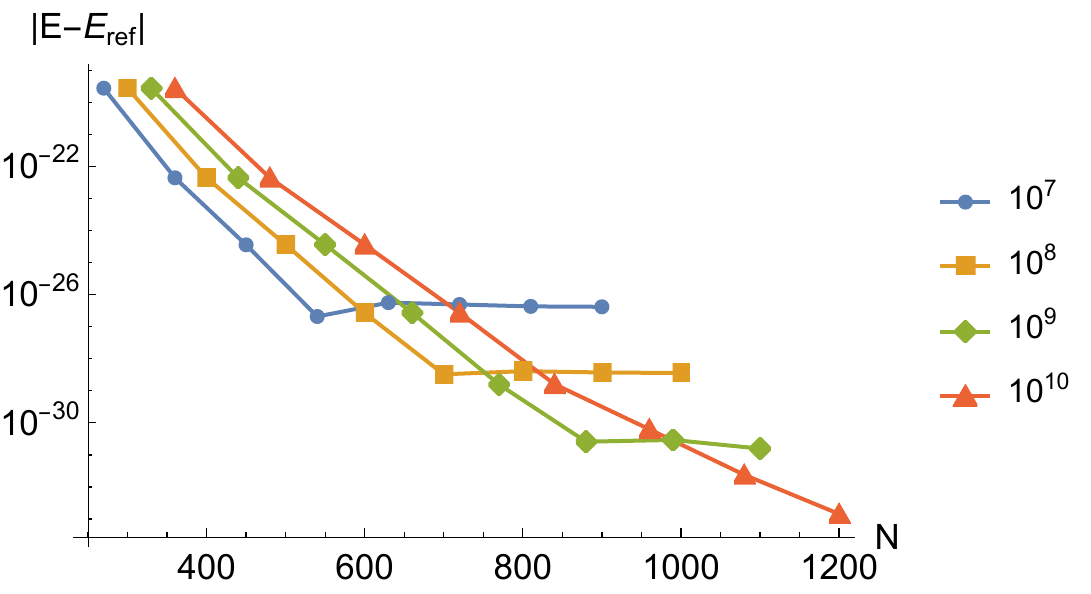}\\
  \caption{Same as Fig.~\ref{conv_e_nkb}, for the DKB basis set.}
  \label{conv_e_dkb}
\end{figure}

Table \ref{tab_r_dkb} shows the convergence of the ground-state ($1s\sigma_g$) energy of the H$_2^+$ molecular ion for an internuclear distance $R=2.0$~a.u. obtained using the DKB approach. Similar data for NKB is given in Table~\ref{tab_r_nkb} in the Appendix~\ref{sec-app-nkb}. In addition, the convergence for both basis sets is shown graphically, using a more extensive set of data with respect to the Tables, in Figs.~\ref{conv_e_nkb} and \ref{conv_e_dkb}. 

\begin{table*}[t]
\begin{center}
\begin{tabular}{@{\hspace{2mm}}l@{\hspace{3mm}}l@{\hspace{3mm}}l@{\hspace{3mm}}l@{\hspace{2mm}}}
\hline\hline
\vrule width0pt height10pt depth4pt
$n_i$ & \multicolumn{1}{c}{$\alpha_{\rm max} = 10^8$} & \multicolumn{1}{c}{$\alpha_{\rm max} = 10^{10}$} \\
\hline
\vrule width0pt height10pt depth4pt
$30$      & $\mathbf{-1.102\,641\,581\,032\,577\,164}\,089\,813\,929\,495$  & $\mathbf{-1.102\,641\,581\,032\,577\,164}\,089\,813\,916\,081\,035$ &\\
$40$      & $\mathbf{-1.102\,641\,581\,032\,577\,164\,118\,1}70\,109\,212$  & $\mathbf{-1.102\,641\,581\,032\,577\,164\,118}\,170\,109\,063\,546$ &\\
$50$      & $\mathbf{-1.102\,641\,581\,032\,577\,164\,118\,12}5\,368\,536$  & $\mathbf{-1.102\,641\,581\,032\,577\,164\,118\,12}5\,368\,576\,888$ &\\
$60$      & $\mathbf{-1.102\,641\,581\,032\,577\,164\,118\,125\,00}2\,656$  & $\mathbf{-1.102\,641\,581\,032\,577\,164\,118\,125\,00}2\,692\,652$ &\\
$70$      & $\mathbf{-1.102\,641\,581\,032\,577\,164\,118\,124\,999\,9}24$  & $\mathbf{-1.102\,641\,581\,032\,577\,164\,118\,124\,999\,9}73\,845$ &\\
$80$      & $\mathbf{-1.102\,641\,581\,032\,577\,164\,118\,124\,999\,9}16$  & $\mathbf{-1.102\,641\,581\,032\,577\,164\,118\,124\,999\,95}8\,261$ &\\
$90$      & $\mathbf{-1.102\,641\,581\,032\,577\,164\,118\,124\,999\,9}20$  & $\mathbf{-1.102\,641\,581\,032\,577\,164\,118\,124\,999\,957\,6}80$ &\\
$100$     & $\mathbf{-1.102\,641\,581\,032\,577\,164\,118\,124\,999\,9}21$  & $\mathbf{-1.102\,641\,581\,032\,577\,164\,118\,124\,999\,957\,65}4$ &\\
\hline
\vrule width0pt height10pt depth4pt
& \multicolumn{1}{c}{$\alpha_{\rm max} = 10^{11}$} &  \\
\hline
\vrule width0pt height10pt depth4pt
$120$     & $-1.102\,641\,581\,032\,577\,164\,118\,124\,999\,957\,656 \, 2$ & \\
\hline
\hline
\end{tabular}
\end{center}
\caption{Convergence of the ground-state energy of H$_2^+$ at $R = 2.0$ with the DKB basis set, using different values of the maximal exponent included in the basis, $\alpha_{\rm max}$. Bold figures are converged. The value $E_{\rm ref}$, which is used to estimate the error $|E - E_{\rm ref}|$ in Figs.~\ref{conv_e_nkb} and \ref{conv_e_dkb}, is given in the last line. Values with $\alpha_{\rm max} \geq 10^{10}$ and $n_i \geq 80$ were obtained using 150 digits of numerical precision.}\label{tab_r_dkb}
\end{table*}

The convergence is studied as a function of two parameters: (i) the maximal value of exponents included in the basis, $\alpha_{\rm max}$, which is varied from $10^{7}$ to $10^{10}$ by keeping the first $p$ subsets, where $p$ lies between 9 and 12; (ii) the number of basis functions $n_i$ in each subset. The basis size is equal to $N = pn_i$ per spinor component.

Similar behaviors are observed in NKB and DKB, with quicker convergence in the DKB case. The precision improves with increasing basis size up to a certain value of $N$, above which it saturates. When the maximal exponent $\alpha_{\rm max}$ is increased, the saturation occurs at higher $N$ and a better precision floor is reached. The existence of this precision floor dependent on $\alpha_{\rm max}$ can be understood by considering that the basis set allows representing the behavior of the wavefunction down to a distance $r \sim 1/\alpha_{\rm max}$ from the nuclei. The scaling of the error on the energy can be estimated in a simplified approach by calculating the contribution to the energy from a sphere of radius $1/\alpha_{\rm max}$ centered on a nucleus, taking into account the short-distance behavior of the wavefunction, $\varphi \sim r^{\gamma - 1}$ with $\gamma = \sqrt{1 - Z^2/c^2}$. One then gets $\Delta E \sim (1/\alpha_{\rm max})^{2 \gamma}$. A power-law fit of our data as a function of $1/\alpha_{\rm max}$ yields exponents of $2.16$ for $Z=1$, whereas $2\gamma \simeq 2.00$, and $1.58$ for $Z = 90$~(see the convergence data in Table~\ref{tab_r2_dkb} in the Appendix~\ref{sec-app-z90}), whereas $2\gamma \simeq 1.51$, in reasonable agreement with the above model.

A phenomenon known as ``prolapse''~\cite{Faegri01,Tatewaki03,Dyall12} is observed for small values of $n_i$, i.e., the energy lies below the exact energy, which is possible because the Dirac energy is not a variational minimum. This behavior disappears at larger basis sizes: the values obtained in the saturation region are always above the exact energy and decrease when $\alpha_{\rm max}$ is increased.

Overall, the NKB and DKB basis sets yield the ground-state energy with 27 and 32 converged digits, respectively. The large improvement with respect to RKB (see~\cite{Nogueira22} and Appendix~\ref{sec-app-rkb}) is consistent with the discussion in~\cite{Nogueira22}, where it was noted that the behavior of the small components $\chi$ in the vicinity of the point-like nuclei is not well represented by the RKB prescription. Our results indicate that the simple exponential basis functions used in NKB improve the description of $\chi$, and that the best description is provided by the more flexible DKB basis set, which includes both the pure exponential behavior and that imposed by kinetic balance. 

The faster convergence of the DKB approach comes at the cost of a higher complexity of the matrix elements, requiring more computation time. Depending on the application, either DKB or NKB may turn out to be the most cost-effective method.

Our final results for the H$_2^+$ ($Z=1$) and Th$_2^{179+}$ ($Z=90$) are given in Table~\ref{tab_r2_pt2} and compared to previous works. The precision decreases at high $Z$ because of the stronger singularity of the wavefunction at the nuclei, which slows down the convergence with respect to $\alpha_{\rm max}$, as discussed above. Despite this, the precision is improved by five orders of magnitude, showing the potential interest of our approach for calculations in heavy quasi-molecules.

\begin{table}[t]
\begin{center}
\begin{tabular}{@{\hspace{2mm}}c@{\hspace{2mm}}l@{\hspace{2mm}}c@{\hspace{2mm}}c@{\hspace{2mm}}}
\hline\hline
$Z$ & \multicolumn{1}{c}{$E$} & $c$ & Ref. \\
\hline
\vrule width0pt height10pt depth4pt
 $1$ & $-1.102\,641\,581\,032\,577\,164\,118\,124\,999\,957\,65$ & $c_a$ & this work \\
     & $-1.102\,641\,581\,032\,577\,164\,118\,1$                 & $c_a$ & \cite{Kullie22} \\
\hline
\vrule width0pt height10pt depth4pt
$90$ & $-9\,504.756\,648\,434\,009\,50$                           & $c_a$ & this work \\
     & $-9\,504.756\,648\,536\,783\,47$                           & $c_b$ & this work \\
     & $-9\,504.756\,648\,531$                                   & $c_b$ & \cite{Tupitsyn14} \\
\hline
\hline
\end{tabular}
\end{center}
\caption{Comparison of the Dirac ground-state energy obtained in this work, using DKB, with previous results. The value of~\cite{Kullie22} is obtained from the more precise value of the relativistic correction given in the added note~\cite{Kullie}. For $Z = 90$, we repeated the calculation with a different value of $c$, $c_b=137.035\,999\,074$ to compare our result with that of Ref.~\cite{Tupitsyn14}. In the results of 'this work', all digits are converged.}\label{tab_r2_pt2}
\end{table}

Beyond the ground-state energy, the diagonalization of the eigenvalue problem gives a numerical representation of the full spectrum of the Dirac Hamiltonian, whose precision can be assessed through the calculation of sum rules, as described in the next section. For illustration, the energies of the first few excited states of H$_2^+$ can be found in Table~\ref{tab_es}. No detailed convergence study was undertaken, but the 21 given digits are converged for all levels. It is worth noting that no spurious states were found in these calculations: after addition of $c^2$ that was subtracted in Eq.~(\ref{eq_dirac}b), the $4N$ eigenvalues split into $2N$ positive eigenvalues, and $2N$ in the negative-energy continuum below $E = -c^2$. However, two spurious states were found for $Z = 90$ for the largest basis sizes ($n_i \ge 90$).

\begin{table*}[t]
\begin{center}
\begin{tabular}{@{\hspace{1mm}}c@{\hspace{2mm}}c@{\hspace{2mm}}l@{\hspace{2mm}}l@{\hspace{2mm}}}
\hline\hline
\vrule width0pt height10pt depth4pt
State & $|j_z|$ & \multicolumn{1}{c}{$E$ (this work)} & \multicolumn{1}{c}{$E$~\cite{Fillion12}} \\
\hline
\vrule width0pt height10pt depth4pt
$2p\sigma_u$ & $1/2$ & %$\mathbf{-0.667\,552\,771\,993\,113\,045\,809\,8}31\,0$  \\
$-0.667\,552\,771\,993\,113\,045\,809$ & $-0.667 \, 552 \, 771 \, 8$ \\
$2p\pi_u$    & $1/2$ & %$\mathbf{-0.428\,781\,160\,212\,631\,303\,442\,061\,152\,112\,697\,860}$
$-0.428\,781\,160\,212\,631\,303\,442$ & $-0.428 \, 781 \, 158 \, 4$\\
$2p\pi_u$    & $3/2$ & %$\mathbf{-0.428\,774\,447\,992\,646\,216\,404\,344\,515\,141\,669\,855\,5}43\,8$  \\
$-0.428\,774\,447\,992\,646\,216\,404$ & \\
$2s\sigma_g$ & $1/2$ & %$\mathbf{-0.360\,871\,070\,577\,597\,640\,901\,330}\,890$  \\
$-0.360\,871\,070\,577\,597\,640\,901$ & $-0.360 \, 871 \, 069 \, 5$ \\
$3p\sigma_u$ & $1/2$ & %$\mathbf{-0.255\,419\,704\,748\,235\,324\,061\,044\,684\,758\,056\,301}$ \\ 
$-0.255\,419\,704\,748\,235\,324\,061$ & $-0.255 \, 419 \, 703 \, 3$ \\
$3d\sigma_g$ & $1/2$ & %$\mathbf{-0.235\,781\,268\,452\,381\,629\,103\,121\,8}665$  \\
$-0.235\,781\,268\,452\,381\,629\,103$ & $-0.235 \, 781 \, 268 \, 1$ \\
$3d\pi_g$    & $1/2$ & %$\mathbf{-0.226\,703\,071\,340\,986\,072\,893\,258\,768\,12}5\,15$  \\
$-0.226\,703\,071\,340\,986\,072\,893$ & $-0.226 \, 703 \, 069 \, 6$ \\
$3d\pi_g$    & $3/2$ & %$\mathbf{-0.226\,701\,493\,971\,348\,876\,268\,993\,933\,111\,021\,498}\,181$  \\
$-0.226\,701\,493\,971\,348\,876\,268$ &\\
\hline
\hline
\end{tabular}
\end{center}
\caption{Energies of the first eight excited states of H$_2^+$ at $R=2.0$. All digits are converged. Note that $\pi$ states give rise to a fine-structure doublet.}\label{tab_es}
\end{table*}

\section{Sum rules} \label{sec-sum-rules}

In this section, we use the eigenvalues, $E_n$, and eigenvectors, $|\psi_n\rangle$, obtained by full diagonalization of the Dirac equation expanded in the NKB basis set [Eq.~(\ref{eq_nkb})] to evaluate the sum rules
\begin{equation} \label{sum_rules}
S_i = \sum_n \left(E_n - E_0\right)^i |\langle\Psi_0|\mathbf{r}|\Psi_n\rangle|^2,
\end{equation}
as done in~\cite{Drake81}. $\textbf{r}$ is the position vector of the electron. $E_0$ and $|\Psi_0\rangle$ are the energy and wavefunction of the ground state, which is an even state with $j_z= \pm 1/2$. Choosing $j_z= 1/2$, it is coupled via the $\textbf{r}$ operator to odd states having $j_z =-1/2$, $1/2$, and $3/2$. The index $n$ therefore runs over all states having these symmetries.

The first values of $S_i$ can be shown to be~\cite{Drake81,Dalgarno69}:
\begin{subequations} \label{sr_exact}
\begin{flalign}
&S_0 = \langle\Psi_0|r^2|\Psi_0\rangle \,,&  \\
&S_1 = 0 \,,&\\
&S_2 = 3c^2.& 
\end{flalign}
\end{subequations}
Comparison of the values of $S_i$ obtained with Eq.~(\ref{sum_rules}) with those of Eq.~(\ref{sr_exact}), which are either exact (for $S_1$ and $S_2$) or can be calculated with high accuracy (for $S_0$), provides a way to evaluate the accuracy of our discrete representation of the Dirac spectrum. This can be viewed as a test of accuracy of the numerical Green function
\begin{equation}
G(z) \simeq \sum_n \frac{| \psi_n \rangle \langle \psi_n |}{E_n - z} \,,
\end{equation}
since the $S_i$ can be written in the form
\begin{equation}
S_i = \langle \psi_0 | \mathbf{r} G(E_0) (H - E_0)^{i+1} \mathbf{r} | \psi_0 \rangle \,.
\end{equation}
Results are shown in Table~\ref{tab_sr}, where $\Delta S_i=S_i^{\rm num}-S_i^{\rm exact}$, with $S_i^{\rm exact}$ given by Eq.~(\ref{sr_exact}). The errors are small and decrease as the basis size is increased. This provides strong evidence that our numerical description of the Dirac spectrum is accurate and complete.

\begin{table}[t]
\begin{center}
\begin{tabular}{@{\hspace{4mm}}c@{\hspace{8mm}}c@{\hspace{4mm}}c@{\hspace{4mm}}c@{\hspace{4mm}}}
\hline\hline
\vrule width0pt height10pt depth4pt
$n_i$ & $-\Delta S_0/S_0$ & $-S_1$ & $\Delta S_2/S_2$ \\
\hline
\vrule width0pt height10pt depth4pt
$30$  &  $5.8\!\times\!10^{-14}$  & $1.4\!\times\!10^{-12}$ & $2.5\!\times\!10^{-8}$ \\
$40$  &  $1.4\!\times\!10^{-16}$  & $2.3\!\times\!10^{-15}$ & $4.8\!\times\!10^{-11}$ \\
$50$  &  $2.8\!\times\!10^{-18}$  & $7.2\!\times\!10^{-18}$ & $3.7\!\times\!10^{-12}$ \\
$60$  &  $1.5\!\times\!10^{-20}$  & $8.5\!\times\!10^{-20}$ & $1.1\!\times\!10^{-14}$ \\
$70$  &  $1.3\!\times\!10^{-22}$  & $5.0\!\times\!10^{-22}$ & $4.3\!\times\!10^{-17}$ \\
$80$  &  $1.0\!\times\!10^{-23}$  & $2.4\!\times\!10^{-23}$ & $9.9\!\times\!10^{-18}$ \\
$90$  &  $2.1\!\times\!10^{-24}$  & $1.8\!\times\!10^{-24}$ & $3.6\!\times\!10^{-19}$ \\
$100$ &  $2.1\!\times\!10^{-24}$  & $7.3\!\times\!10^{-26}$ & $2.3\!\times\!10^{-21}$ \\
\hline\hline
\end{tabular}
\end{center}
\caption{Sum rules (see Eq.~(\ref{sum_rules})) for the ground state of H$_2^+$ at $R = 2.0$, using the NKB basis set with $\alpha_{\rm max} = 10^8$.}\label{tab_sr}
\end{table}

\section{Conclusion}

We have shown that the two-center Dirac equation for H$_2^+$ can be solved to essentially arbitrary accuracy using an expansion in a basis set of pure exponential basis functions and multiple-precision arithmetic. Several kinetic balance conditions were compared; the DKB scheme~\cite{Shabaev04} was found to yield the fastest convergence. Alternatively, a pure exponential basis without any kinetic balance condition (NKB) can be used, when the slower convergence is to some extent counterbalanced by simpler calculation of matrix elements. Finally, the calculation of sum rules gave evidence that the full diagonalization of the Dirac Hamiltonian provides an accurate representation of the Green function. This method appears to be a promising tool for high-precision relativistic calculations of molecular properties such as QED corrections, in low-$Z$ but also in high-$Z$ systems.

\textbf{Acknowledgements.} We thank L. Hilico and V. I. Korobov for useful comments on the manuscript. Support of the French Agence Nationale de la Recherche (ANR) under Grant No. ANR-19-CE30-0029 is acknowledged.

\appendix

\section{Restricted kinetic balance} \label{sec-app-rkb}

The RKB prescription consists in imposing the following relationship between the basis functions of the large and small components~\cite{Stanton84,Dyall90}:
\begin{equation} \label{eq_basis_rkb}
f_{\mu} = \frac{1}{2c} \boldsymbol{\sigma}\mathbf{p} \, g_{\mu} \,.
\end{equation}

\subsection{Matrix form of the Dirac equation}

Injecting~(\ref{eq_basis_rkb}) into the Dirac equation, Eq.~(\ref{eq_dirac}), leads to the following matrix form of the Dirac equation~\cite{Sun11}:
\begin{equation} \label{eq_rkb}
\begin{aligned}
\begin{pmatrix}
\mathbf{V} & 2c\mathbf{T} \\
2c\mathbf{T} & \mathbf{W} -4c^2\mathbf{T}
\end{pmatrix}
&\,\begin{pmatrix}
\mathbf{A} \\
\mathbf{B}
\end{pmatrix} \\
& = E \begin{pmatrix}
\mathbf{S} & 0 \\
0   & 2\mathbf{T}
\end{pmatrix}
\begin{pmatrix}
\mathbf{A} \\
\mathbf{B}
\end{pmatrix} ,
\end{aligned}
\end{equation}
where the matrix elements are
\begin{equation} \label{eq_elem_rkb}
\begin{aligned}
 &\,\mathbf{V}_{\mu\nu}=\left\langle g_{\mu}|V|g_{\nu}\right\rangle,
 \mathbf{T}_{\mu\nu}=\left\langle g_{\mu}\left|p^2/2\right|g_{\nu}\right\rangle, \\
 & \mathbf{S}_{\mu\nu}=\left\langle g_{\mu}|g_{\nu}\right\rangle, 
 \mathbf{W}_{\mu\nu}=\left\langle g_{\mu}|\boldsymbol{\sigma}\textbf{\text{p}}V\boldsymbol{\sigma}\textbf{\text{p}}|g_{\nu}\right\rangle.
\end{aligned}
\end{equation}

\subsection{Numerical results}

We implemented Eq.~(\ref{eq_rkb}) using the exponential basis functions described in Sec.~\ref{sec-method}. Our results are presented in Table~\ref{tab_r_rkb}. The dependence of the energy on $n_i$ is very close to what was obtained in~\cite{Nogueira22} using a similar RKB basis set and an iterative method (see Table II in that reference). This confirms the equivalence between the direct resolution of the four-component eigenvalue problem, Eq.~(\ref{eq_rkb}), and the method of~\cite{Nogueira22,Kutzelnigg89} based on iterated resolution of a two-component linear system. Results obtained with $\alpha_{\rm max} = 10^8$ and $10^{10}$ are essentially identical, showing that the precision is only limited by the slow convergence with respect to $n_i$. Extrapolation to $n_i \to \infty$ would yield the same value of the ground-state energy as that published in~\cite{Nogueira22}, with an uncertainty of about $10^{-20}$~a.u. However, we do not pursue this analysis here as both the NKB and DKB basis sets provide much faster convergence and more accurate results, as described in Sec.~\ref{sec-results}.

\begin{table*}[h!]
\begin{center}
\begin{tabular}{@{\hspace{2mm}}l@{\hspace{3mm}}l@{\hspace{3mm}}l@{\hspace{3mm}}l@{\hspace{2mm}}}
\hline\hline
\vrule width0pt height10pt depth4pt
$n_i$ & \multicolumn{1}{c}{$\alpha_{\rm max} = 10^8$} & \multicolumn{1}{c}{$\alpha_{\rm max} = 10^{10}$} \\
\hline
\vrule width0pt height10pt depth4pt
$30$      & $\mathbf{-1.102\,641\,581\,032\,577\,164}\,817\,576$ & $\mathbf{-1.102\,641\,581\,032\,577\,164}\,817\,577$ &\\
$40$      & $\mathbf{-1.102\,641\,581\,032\,577\,164}\,463\,630$ & $\mathbf{-1.102\,641\,581\,032\,577\,164}\,463\,628$ &\\
$50$      & $\mathbf{-1.102\,641\,581\,032\,577\,164}\,238\,103$ & $\mathbf{-1.102\,641\,581\,032\,577\,164}\,238\,103$ &\\
$60$      & $\mathbf{-1.102\,641\,581\,032\,577\,164\,1}50\,087$ & $\mathbf{-1.102\,641\,581\,032\,577\,164\,1}50\,087$ &\\
$70$      & $\mathbf{-1.102\,641\,581\,032\,577\,164\,1}41\,481$ & $\mathbf{-1.102\,641\,581\,032\,577\,164\,1}41\,481$ &\\
$80$      & $\mathbf{-1.102\,641\,581\,032\,577\,164\,1}33\,131$ & $\mathbf{-1.102\,641\,581\,032\,577\,164\,1}33\,131$ &\\
$90$      & $\mathbf{-1.102\,641\,581\,032\,577\,164\,1}30\,716$ & $\mathbf{-1.102\,641\,581\,032\,577\,164\,1}30\,716$ &\\
$100$     & $\mathbf{-1.102\,641\,581\,032\,577\,164\,1}26\,142$ & $\mathbf{-1.102\,641\,581\,032\,577\,164\,1}26\,142$ &\\
\hline
\hline
\end{tabular}
\end{center}
\caption{Convergence of the ground-state energy of H$_2^+$ at $R = 2.0$ with the RKB basis set, using different values of the maximal exponent included in the basis, $\alpha_{\rm max}$. Bold figures are converged.}\label{tab_r_rkb}
\end{table*}

\section{No kinetic balance} \label{sec-app-nkb}

Table~\ref{tab_r_nkb} shows our numerical results for the ground-state ($1s\sigma_g$) energy of the H$_2^+$ molecular ion ($Z = 1$) for an internuclear distance $R=2.0$~a.u. obtained using the NKB approach. The convergence is slower than with DKB (see Table~\ref{tab_r_dkb}), but NKB still yields 27-digit accuracy for the largest basis size tested here.

\begin{table*}[t]
\begin{center}
\begin{tabular}{@{\hspace{2mm}}l@{\hspace{3mm}}l@{\hspace{3mm}}l@{\hspace{3mm}}l@{\hspace{2mm}}}
\hline\hline
\vrule width0pt height10pt depth4pt
$n_i$ & \multicolumn{1}{c}{$\alpha_{\rm max} = 10^8$} & \multicolumn{1}{c}{$\alpha_{\rm max} = 10^{10}$} \\
\hline
\vrule width0pt height10pt depth4pt
$30$      & $\mathbf{-1.102\,641\,581\,032\,5}99\,064\,556\,576\,400\,842$ & $\mathbf{-1.102\,641\,581\,032\,5}99\,064\,558\,575\,251\,726$ &\\
$40$      & $\mathbf{-1.102\,641\,581\,032\,577}\,206\,904\,005\,189\,779$ & $\mathbf{-1.102\,641\,581\,032\,577}\,206\,903\,909\,324\,324$ &\\
$50$      & $\mathbf{-1.102\,641\,581\,032\,577\,164}\,396\,302\,629\,401$ & $\mathbf{-1.102\,641\,581\,032\,577\,164}\,396\,294\,032\,166$ &\\
$60$      & $\mathbf{-1.102\,641\,581\,032\,577\,164\,1}20\,584\,096\,424$ & $\mathbf{-1.102\,641\,581\,032\,577\,164\,1}20\,583\,064\,101$ &\\
$70$      & $\mathbf{-1.102\,641\,581\,032\,577\,164\,118\,1}41\,273\,724$ & $\mathbf{-1.102\,641\,581\,032\,577\,164\,118\,1}42\,857\,619$ &\\
$80$      & $\mathbf{-1.102\,641\,581\,032\,577\,164\,118\,12}3\,938\,353$ & $\mathbf{-1.102\,641\,581\,032\,577\,164\,118\,125}\,507\,671$ &\\
$90$      & $\mathbf{-1.102\,641\,581\,032\,577\,164\,118\,12}3\,415\,736$ & $\mathbf{-1.102\,641\,581\,032\,577\,164\,118\,125\,0}19\,588$ &\\
$100$     & $\mathbf{-1.102\,641\,581\,032\,577\,164\,118\,12}3\,227\,381$ & $\mathbf{-1.102\,641\,581\,032\,577\,164\,118\,125\,00}0\,808$ &\\
\hline
\hline
\end{tabular}
\end{center}
\caption{Same as Table~\ref{tab_r_rkb}, using the NKB basis set.}\label{tab_r_nkb}
\end{table*}

\section{Ground-state energy of $\mbox{Th}_2^{179+}$} \label{sec-app-z90}

In order to study the applicability of our approach to strongly bound (high-$Z$) systems, we calculated the ground-state energy of the Th$_2^{179+}$ molecule ($Z = 90$) at $R = 2.0/Z$~a.u. using DKB. The basis set is obtained by multiplying by $Z$ the bounds of the intervals in which the exponents $\alpha_i, \beta_i$ are generated. Our results are shown in Table~\ref{tab_r2_dkb}. The convergence is much slower than for $Z = 1$ (compare with~Table \ref{tab_r_dkb}), but we were still able to obtain 18 converged digits, which represents an improvement by 5 orders of magnitude with respect to Ref.~\cite{Tupitsyn14}.

\begin{table*}[h!]
\begin{center}
\begin{tabular}{@{\hspace{2mm}}l@{\hspace{3mm}}l@{\hspace{3mm}}l@{\hspace{3mm}}l@{\hspace{2mm}}}
\hline\hline
\vrule width0pt height10pt depth4pt
$n_i$ & \multicolumn{1}{c}{$\alpha_{\rm max}/Z = 10^8$} & \multicolumn{1}{c}{$\alpha_{\rm max}/Z = 10^{10}$} \\
\hline
\vrule width0pt height10pt depth4pt
$60$      & $\mathbf{-9\,504.756\,648\,434\,00}7\,951\,288$ & $\mathbf{-9\,504.756\,648\,434\,009\,49}9\,551$ &\\
$70$      & $\mathbf{-9\,504.756\,648\,434\,00}7\,761\,385$ & $\mathbf{-9\,504.756\,648\,434\,009\,49}9\,550$ &\\
$80$      & $\mathbf{-9\,504.756\,648\,434\,00}7\,970\,559$ & $\mathbf{-9\,504.756\,648\,434\,009\,49}9\,570$ &\\
$90$      & $\mathbf{-9\,504.756\,648\,434\,00}8\,102\,338$ & $\mathbf{-9\,504.756\,648\,434\,009\,49}9\,639$ &\\
$100$     & $\mathbf{-9\,504.756\,648\,434\,00}8\,162\,438$ & $\mathbf{-9\,504.756\,648\,434\,009\,49}9\,723$ &\\
\hline
\vrule width0pt height10pt depth4pt
& \multicolumn{1}{c}{$\alpha_{\rm max}/Z = 10^{11}$} &  \\
\hline
\vrule width0pt height10pt depth4pt
$120$     & $-9\,504.756\,648\,434\,009\,500\,732$ & \\
\hline
\hline
\end{tabular}
\end{center}
\caption{Convergence of the ground-state energy of Th$_2^{179+}$ ($Z = 90$) at $R = 2.0/Z$ with the DKB basis set. Values with $\alpha_{\rm max}/Z = 10^{10}$ and $n_i \geq 90$ were obtained using 150 digits of numerical precision, while the value with $\alpha_{\rm max}/Z = 10^{11}$ and $n_i = 120$ was obtained using 300 digits of numerical precision.}\label{tab_r2_dkb}
\end{table*}

\end{document}